\documentclass[prl,aps,showpacs,reprint,superscriptaddress,amsmath,amssymb,floatfix,10pt]{revtex4-1} 
\usepackage{amsmath,amssymb,graphicx}
\usepackage[colorlinks=true, citecolor=blue, linkcolor=blue, urlcolor=blue]{hyperref}
\usepackage{array} 

\begin{document}

\title{Superresolution via Structured Illumination Quantum Correlation Microscopy (SIQCM)}

\author{Anton Classen}
\affiliation{Institut f\"ur Optik, Information und Photonik, Universit\"at Erlangen-N\"urnberg,
	91058 Erlangen, Germany}
\affiliation{Erlangen Graduate School in Advanced Optical Technologies (SAOT), Universit\"at Erlangen-N\"urnberg,
	91052 Erlangen, Germany}

\author{Joachim von Zanthier}
\affiliation{Institut f\"ur Optik, Information und Photonik, Universit\"at Erlangen-N\"urnberg,
	91058 Erlangen, Germany}
\affiliation{Erlangen Graduate School in Advanced Optical Technologies (SAOT), Universit\"at Erlangen-N\"urnberg,
	91052 Erlangen, Germany}

\author{Marlan O. Scully}
\affiliation{Texas A\&M University, College Station, Texas 77843-4242, USA}
\affiliation{Princeton University, Princeton, New Jersey 08544, USA}
\affiliation{Baylor University, Waco, Texas 76798, USA}

\author{Girish S. Agarwal}
\affiliation{Texas A\&M University, College Station, Texas 77843-4242, USA}
\affiliation{Department of Physics, Oklahoma State University, Stillwater, Oklahoma 74078, USA}

\begin{abstract}
We propose to use intensity correlation microscopy in combination with structured illumination to image quantum emitters that exhibit antibunching with a spatial resolution reaching far beyond the Rayleigh limit. 
Combining intensity measurements and intensity auto correlations up to order $m$ creates an effective PSF with FWHM shrunk by the factor $\sqrt{m}$. Structured Illumination microscopy on the other hand introduces a resolution improvement of factor 2 by use of the principle of moir\'e fringes. Here, we show that for linear low-intensity excitation and linear optical detection the simultaneous use of both techniques leads to an in theory unlimited resolution power with the improvement scaling favorably as $m + \sqrt{m}$ in dependence of the correlation order $m$. Hence, yielding this technique to be of interest in microscopy for imaging a variety of samples including biological ones. We present the underlying theory and simulations demonstrating the highly increased spatial superresolution, and point out requirements for an experimental implementation.
\end{abstract}

\maketitle 

\section{Introduction}

Superresolution optical far-field microscopy has undergone a tremendous evolution since roughly two decades ago it was shown that the classical resolution limit \cite{Abbe1873,Rayleigh1879-2} posed by diffraction can be overcome \cite{Hell1994,Hell2015}, resulting in the development of a large variety of methods achieving superresolution. One group of methods relies on stimulated ground or excited state depletion and a non-linear response of fluorescence markers to given excitation intensities to deterministically engineer the effective excitation point spread function (PSF) \cite{Hell1994,Hell2015,Hell1995,Hell2005}. Other methods stochastically localize single photoswitchable molecules with an accuracy of a few ten nanometers via centroid fitting of the PSF \cite{Moerner1997,Mason2006,Xiaowei2006,Betzig2006}. Another branch of methods makes use of higher-order intensity cross correlations in the Fourier plane \cite{Thiel2007,Oppel2012a,Classen2016} or auto correlations in the image plane of a microscope \cite{Dertinger2009,Schwartz2012,Schwartz2013,Genovese2014a}. For the latter group of correlation microscopy techniques (CM), either super-poissonian bunched light emission due to statistical fluctuations \cite{Dertinger2009} or sub-poissonian anti-bunched light emission of fluorescence markers can be used to enhance the resolution, both in widefield \cite{Schwartz2013} and confocal microscopy \cite{Genovese2014a}. Finally, structured illumination microscopy (SIM) leads to a doubled resolution by use of the principle of moir\'e fringes and linear wave optics \cite{Gustafsson2000}, and its non-linear derivative saturated SIM (SSIM) leads to further improvements and in principle unlimited resolution, though at the cost of necessitating high intensities \cite{Gustafsson2005}. Other derivatives combine SIM with the third-order process of CARS or with graphene plasmons to access more higher spatial frequency information than ordinary SIM \cite{Rubinsztein2010,Lee2014,Zubairy2014}. Note that sub-wavelength phenomena can also be found in other fields of physics, for example in sub-wavelength atom localization due to the non-linear behavior of coherent population trapping (CPT) \cite{Agarwal2006a,Yavuz2013,Yavuz2015} and sub-wavelength lithography via Rabi-oscillations \cite{Zubairy2008,Zubairy2010}. CPT was also proposed to highly increase the resolution in a microscopy themed derivative \cite{Kapale2010}. 

Here we report on a novel superresolution method that relies on intensity correlation measurements in the image plane of a microscope in combination with 
structured illumination to image fluorophores that exhibit anti-bunching. We therefore term it Structured Illumination Quantum Correlation Microscopy (SIQCM). Linear low-intensity excitation and linear detection suffice such that the technique holds promise to highly enhance the resolution in biological imaging. Detrimental effects due to high intensities that are required by many superresolution techniques, leading to phototoxicity and photobleaching in fluorophores, do not arise. We demonstrate that already very low correlation orders $m$ provide highly enhanced  superresolution, that scales favorably as $m+\sqrt{m}$. The present manuscript focuses on the highly enhanced lateral resolution 
using a simple widefield microscopic setup, however one can easily extend the scheme as CM as well as SIM each on their own already provide optical sectioning capability for 3D imaging \cite{Dertinger2012,Schwartz2013,Gustafsson2008}.

The technique makes use of $m$th-order correlations and antibunched photon emission, inherently present in most common fluorophores, even at room temperature \cite{Basche1992,Moerner2000,Grangier2000,Buratto2000}. Hence, the required quantum emitters are already broadly in use in fluorescence microscopy. Furthermore, SIM is a well established technique in biological imaging with commercial microscopes, attaining the theoretically predicted resolution enhancement, being widely spread.

\section{Theory}

Let $h(\mathbf{r})$ be the PSF of a given microscope, where $\mathbf{r}$ denotes the position in the image plane and $H(\mathbf{k}) \equiv FT\{ h(\mathbf{r}) \}$ is the corresponding optical transfer function (OTF) obtained by Fourier transform ($FT$), where $\mathbf{k}$ denotes the spatial frequency in reciprocal space. Later, in the $m$th-order correlation microscopy signals $\text{CM}_{m}$ the effective PSF reads $h_m(\mathbf{r}) \equiv (h(\mathbf{r}))^m$ and its corresponding OTF shall be defined as $H_m(\mathbf{k})$. In general $h_m(\mathbf{r})$ gets narrower for increasing correlation order $m$ and its full width half maximum (FWHM) approximately scales as $1/\sqrt{m}$. Vice versa, the observable region in reciprocal space is increased for $H_m(\mathbf{k})$ by $\sqrt{m}$. Microscopes usually possess the circularly symmetric Airy disk $(2J_1(r)/r)^2$ with $ r = |\mathbf{r}|$ as PSF \cite{BornWolf1999}, what allows to resolve individual incoherent emitters as individual sources of radiation as long as their separation $d$ is at least on the order of $d \geq \lambda/2$ or more precisely $d \geq 0.61\,\lambda/\mathcal{A}$ \cite{Abbe1873,Rayleigh1879-2}. We denote the Rayleigh limit as $d_{\text{R}} \equiv 0.61\,\lambda/\mathcal{A}$, with $\mathcal{A}$ the numerical aperture of the microscope objective and $\lambda$ the wavelength of the emitted fluorescence light. W.l.o.g we assume a magnification of one (or rather minus one) throughout our theoretical treatment such that the coordinates in the object and image plane $\mathbf{R}$ and $\mathbf{r}$, respectively,  can be regarded as equal, i.e. $\mathbf{R} \equiv \mathbf{r}$.

To measure fluorescence photons in the image plane the fluorophores in the object plane need to be driven by an excitation light field. In classical linear optics the fluorophores respond linearly to a given excitation intensity $I_0$. 
Treating the fluorophores quantum mechanically as a two-level system with ground $|g\rangle$ and excited state $|e\rangle$, however, this is only the case for intensities $ I_0 \ll I_\text{sat}$, where the saturation intensity $I_\text{sat} \propto \frac{1}{\tau_l^2} $ depends on the exited states lifetime $\tau_l$. The general expression for the intensity emitted by a two-level system driven by a given excitation intensity (in units of the $I_\text{sat}$) reads \cite{Agarwal2012}
\begin{equation}
I \propto \frac{1}{2} \frac{I_0}{I_\text{sat}+ I _0} \, .
\label{eq:int}
\end{equation}
In ordinary classical microscopy with fluorophores that possess lifetimes on the order of a few ns or below intensities usually remain in the linear regime. To induce non-linear responses, e.g. required by STED microscopy \cite{Hell1994} or SSIM \cite{Gustafsson2005}, very high intensities are necessary that are accompanied by detrimental effects to biological imaging. 
In contrast, our approach contents with low intensity and linear response of fluorophores to achieve highly increased superresolution.
 

Let us first assume a continuous and spatially uniform excitation illumination in the object plane with intensity $I_{\text{str}}(\mathbf{r},t)= I_0$ and the fluorophore density distribution $n(\mathbf{r}) \propto \sum_{i=1}^N \delta(\mathbf{r}-\mathbf{r}_i)$ to be comprised of individual point-like sources at positions $\mathbf{r}_i$ that emit statistically independent, i.e. incoherent radiation. Note that we can also assign (relative) weights to the independent emitters in case their photon emission rates differ. Differences in (relative) emission rates would be enhanced in the (higher-order) intensity auto correlations. However, usually fluorophores emit sufficiently uniform and our technique does not require very high correlation orders to achieve highly enhanced superresolution, in contrast to SOFI \cite{Dertinger2009}. Further, in SOFI this problem is resolved by using balanced cumulants \cite{Geissbuehler2012} and our higher order correlation signals can be adapted accordingly. Therefore, and to keep the analysis illustrative we consider uniform emission rates here.

Considering a linear response, i.e. $I_0 \ll I_\text{sat}$, the intensity in the image plane reads 
\begin{equation}
I(\mathbf{r}) = \langle \hat{E}^{(-)}(\mathbf{r}) \hat{E}^{(+)}(\mathbf{r}) \rangle \propto \sum_{i=1}^N h(\mathbf{r}-\mathbf{r}_i) \, ,
\label{eq:1}
\end{equation}
where $\hat{E}^{(+)}(\mathbf{r})  \propto \sum_{i} (2J_1(|\mathbf{r}-\mathbf{r}_i|)/|\mathbf{r}-\mathbf{r}_i|) \,  e^{i \phi_i} \, \hat{\sigma}_i^{-}$ is the positive frequency part of the electric field operator and $\hat{\sigma}_i^{-}$ is the lowering operator acting on the fluorophore at $\mathbf{r}_i$, which can be approximated by a two-level system with ground and excited states $|g_i \rangle$ and $|e_i \rangle$. The phases $\phi_i$ are varying randomly and independently on time scales larger than the excited states lifetime $\tau_l$ and introduce the incoherence as the expectation value $ \langle e^{i \phi_i}  e^{- i \phi_j}  \rangle = 0$ for $i \neq j$. Note that the intensity $I(\mathbf{r}) \equiv G^{(1)}(\mathbf{r})$ can be recognized as Glauber's first-order equal-time intensity correlation function $G^{(1)}(\mathbf{r}_1,t_1;\mathbf{r}_2=\mathbf{r}_1,t_2=t_1) = \langle \hat{E}^{(-)}(\mathbf{r}_1,t_1) \hat{E}^{(+)}(\mathbf{r}_2,t_2) \rangle$, assuming an ergodic system \cite{Glauber1963-2}.

\begin{figure}%
\centering
\includegraphics[width=1.0 \linewidth]{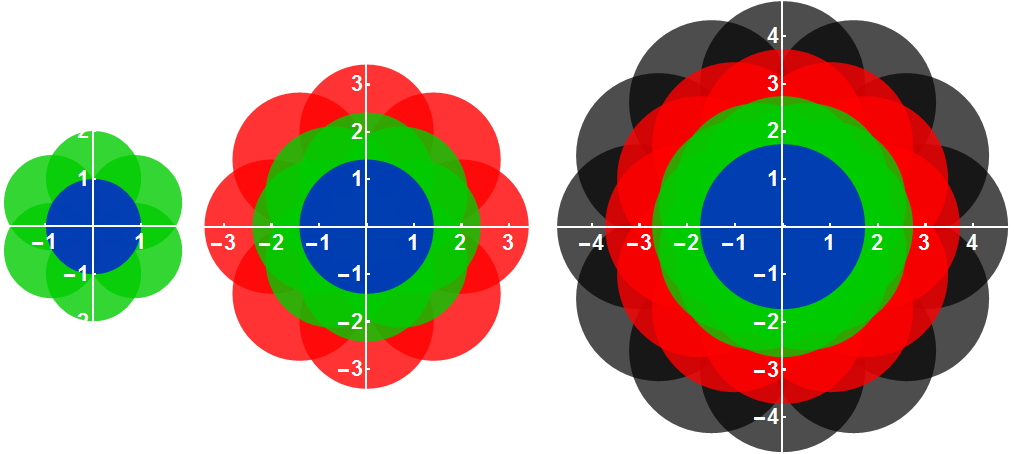}%
\caption{Comparison of the observable region in reciprocal space provided by ordinary SIM and SIQCM for second order and third order.}
\label{fig:SIM+ABM}%
\end{figure}

Taking the square $(G^{(1)}(\mathbf{r}))^2 =  \sum_{i=1}^N (h(\mathbf{r}-\mathbf{r}_i))^2 +  \sum_{i\neq j}^N h(\mathbf{r}-\mathbf{r}_i) h (\mathbf{r}-\mathbf{r}_j)$ we obtain an incoherent sum of narrowed PSFs $h_2(\mathbf{r}-\mathbf{r}_i)$, however in addition also the detrimental cross terms. These cross terms can be removed by subtracting the second-order intensity auto correlation function $ G^{(2)}(\mathbf{r}) \equiv  G^{(2)}(\mathbf{r},\mathbf{r})  = \langle \hat{E}^{(-)}(\mathbf{r}) \hat{E}^{(-)}(\mathbf{r}) \hat{E}^{(+)}(\mathbf{r}) \hat{E}^{(+)}(\mathbf{r}) \rangle \propto 2 \sum_{i\neq j}^N h(\mathbf{r}-\mathbf{r}_i) h(\mathbf{r}-\mathbf{r}_j) $. Here, the squared terms $h_2(\mathbf{r})$ vanish as each two-level system can emit only one photon simultaneously, that is $\langle \hat{\sigma}_i^{+}  \hat{\sigma}_i^{+} \hat{\sigma}_i^{-} \hat{\sigma}_i^{-}  \rangle = 0$. Subtracting the signals we obtain
\begin{equation}
\text{CM}_2(\mathbf{r}) = \left(G^{(1)}(\mathbf{r})\right)^2 - \frac{1}{2} G^{(2)}(\mathbf{r}) = \sum_{i=1}^N h_2(\mathbf{r}-\mathbf{r}_i) \, ,
\label{eq:2}
\end{equation}
what is the sought-after anti-bunching $\text{CM}_2$ signal \cite{Schwartz2012,Schwartz2013,Genovese2014a}. Higher-order $\text{CM}_m$ signals are derived analogously taking into account higher-order correlation functions up to $ G^{(m)}(\mathbf{r})$. The resolution enhancement of this signal moderately scales as $\sqrt{m}$ with the correlation order, what is also illustrated in reciprocal space by the central (blue) circles in Fig.\,\ref{fig:SIM+ABM} that define the observable regions for ordinary intensity measurements, $\text{CM}_2$ and $\text{CM}_3$ (from left to right).

Now, considering a two-dimensional structured illumination $I_{\text{str}}(\mathbf{r},t)= I_0 [  \frac{1}{2} + \frac{1}{2} \cos (\mathbf{k}_0 \mathbf{r} + \varphi ) ]$, a linear response of the fluorophores and ordinary intensity measurements one obtains a doubled resolution by the principle of moir\'e fringes. The illumination pattern and the investigated sample produce beat patterns in the object and the image plane such that initially unobservable spatial frequencies in reciprocal space are shifted by the amount $k_0= | \mathbf{k}_0 | = \sqrt{k_x^2 + k_y^2}$\; into the observable region and thus can be accessed (cf. left side in Fig.\,\ref{fig:SIM+ABM}). In general it is useful to define $I_{\text{str}}=I_{\text{str}}(\mathbf{r},\alpha,\varphi)$, where $\alpha = \tan (k_y/k_x)$ is the orientation and $\varphi$ is the adjustable phase of the pattern. Note that larger $k_0$ effectively enlarge the observable region in reciprocal space and thus the resolution by a higher amount, however $k_0$ is limited by diffraction and the given numerical aperture $\mathcal{A}$ of the microscope objective. Hence, by use of far field wave optics infinitesimally dense fringe spacings in the source plane can not be produced. Following the derivation for Eq.\,(\ref{eq:1}) with adjusted $I_{\text{str}}(\mathbf{r})$, the resulting signal reads (see also Ref. \cite{Gustafsson2000})
\begin{equation}
G^{(1)}(\mathbf{r}) = \sum_{i=1}^N h(\mathbf{r}-\mathbf{r}_i)  \cdot I_{\text{str}}(\mathbf{r}_i,\alpha,\varphi) \, .
\label{eq:3}
\end{equation} 
Rewriting this expression into $h(\mathbf{r}) \ast \left[ n(\mathbf{r}) \cdot I_{\text{str}}(\mathbf{r},\alpha,\varphi) \right]$ and taking the Fourier transform yields
\begin{equation}
\begin{aligned}
 & FT \left\{ h(\mathbf{r}) \ast \left[ n(\mathbf{r}) \cdot I_{\text{str}}(\mathbf{r},\varphi,\alpha) \right] \right\}   \\ & = H(\mathbf{k}) \cdot   \left[  \frac{1}{2} \,\tilde{n}(\mathbf{k})+ \frac{1}{4} e^{i \varphi } \,\tilde{n}(\mathbf{k} -  \mathbf{k}_0) + \frac{1}{4}  e^{-i \varphi} \,  \tilde{n}(\mathbf{k} +  \mathbf{k}_0) \right] \, ,
\end{aligned}
\label{eq:4}
\end{equation}
where we used the convolution theorem and the identity $FT \{ e^{ik_0 r} g(r) \} = \tilde{g}(k-k_0)$. The density in reciprocal space is denoted by $\tilde{n}(\mathbf{k})$ which arises together with its shifted versions, offset by $\pm \mathbf{k}_0 $. One image does not allow to separate the three individual components, such that three images with three different phases $\varphi = 0,\frac{2\pi}{3}, \frac{4\pi}{3}$ are required, creating the linear system $A \vec{n} = \vec{G}$, where the matrix $A$ describes the resulting system and $\vec{n}$ denotes a vector with entries $\tilde{n}(\mathbf{k})$, $\tilde{n}(\mathbf{k} -  \mathbf{k}_0)$ and $\tilde{n}(\mathbf{k} +  \mathbf{k}_0)$. The vector $\vec{G}$ on the right hand side of the system possesses the entries $\tilde{I}(\mathbf{k},\alpha,0)$, $\tilde{I}(\mathbf{k},\alpha,\frac{2\pi}{3})$ and $\tilde{I}(\mathbf{k},\alpha,\frac{4\pi}{3})$ which represent the Fourier transforms of the (experimentally) measured data. The system is solved by applying the inverse matrix $ \vec{n} = A^{-1} \vec{G}$. To sufficiently cover the enlarged area in reciprocal space it is necessary to chose at least three orientations $\alpha = 0,\frac{1\pi}{3}, \frac{2\pi}{3}$ (cf. left side in Fig.\,\ref{fig:SIM+ABM}) resulting in a total of 9 measurements. 

Taking a non-linear fluorophore response into account higher harmonics of $\cos (\mathbf{k}_0 \mathbf{r} + \varphi )$ arise enabling access to higher spatial frequencies in reciprocal space via SSIM. The arising higher harmonics can be read out easily when plugging $I_0 \cos (\mathbf{k}_0 \mathbf{r} )$ into Eq.\,(\ref{eq:int}) as the excitation illumination and compiling the Fourier cosine series which reads
\begin{equation}
I \propto \sum_{n} b_n  \,\cos (n \cdot \mathbf{k}_0 \mathbf{r} ) \, .
\label{eq:Nint}
\end{equation}
However, this comes at the cost of necessitating high intensities that lead to phototoxicity and photobleaching in most biological samples and fluorophores. Furthermore, the Fourier coefficients $b_n$ rapidly decrease with increasing $n$, such that only a limited number of higher harmonics surpasses noise  inherently present in every (experimental) signal. Another drawback is the necessity for a very high number of images as each higher harmonic requires two additional phases $\varphi$ and more orientations $\alpha$ are needed to cover the enlarged observable region in reciprocal space \cite{Gustafsson2005}.

\begin{figure}[ht]%
\centering
\includegraphics[width=1.0 \linewidth]{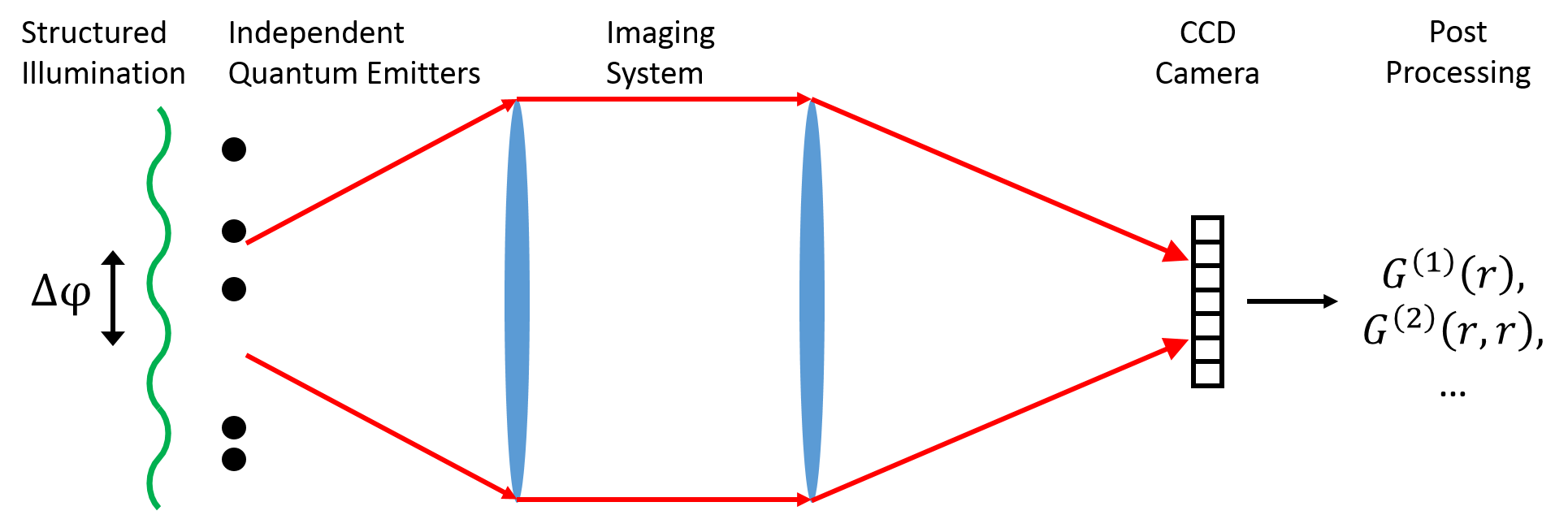}
\caption{Schematic setup to combine SIM and CM to obtain the new SIQCM technique.}
\label{fig:sketch}%
\end{figure}

Our new approach combines the strength of both methods to enhance the already superresolving signals tremendously within the linear low-intensity regime. A schematic sketch is displayed in Fig.\,\ref{fig:sketch}. Considering linear SI $I_{\text{str}}(\mathbf{r},\alpha,\varphi)$ and the $\text{CM}_2$ signal we obtain the SIQCM signal
\begin{equation}
\begin{aligned}
\text{SIQCM}_2(\mathbf{r}) &=  \sum_{i=1}^N h_2(\mathbf{r}-\mathbf{r}_i)  \cdot \left(I_{\text{str}}(\mathbf{r}_i,\alpha,\varphi)\right)^2   \\ 
& = h_2(\mathbf{r}) \ast \left[ n(\mathbf{r}) \cdot\left( I_{\text{str}}(\mathbf{r},\varphi,\alpha) \right)^2 \right] \, ,
\end{aligned}
\label{eq:5}
\end{equation}
and by Fourier transform
\begin{equation}
\begin{aligned}
 H_2(\mathbf{k}) \cdot  & \left[  \frac{1}{2} \,\tilde{n}(\mathbf{k})  + \frac{1}{4} e^{i \varphi } \,\tilde{n}(\mathbf{k} -  \mathbf{k}_0) + \frac{1}{4}  e^{-i \varphi} \,  \tilde{n}(\mathbf{k} +  \mathbf{k}_0) \right. \\
&  \left. + \frac{1}{16} e^{2i \varphi } \,\tilde{n}(\mathbf{k} - 2 \mathbf{k}_0) + \frac{1}{16}  e^{-2i \varphi} \,  \tilde{n}(\mathbf{k} + 2 \mathbf{k}_0) \right] \, ,
\end{aligned}
\label{eq:6}
\end{equation}
where the Fourier components corresponding to the first higher harmonic arise and the individual disks in Fourier space, governed by $H_2(\mathbf{k})$, are enlarged by $\approx \sqrt{2}$, leading to an overall resolution improvement of $2+\sqrt{2} \approx 3.41$. Eq.\,(\ref{eq:6}) is depicted in the middle of Fig.\,\ref{fig:SIM+ABM}, where due to the additional higher harmonic images with five different phases $\varphi =  0,\frac{2\pi}{5}, \frac{4\pi}{5},\frac{6\pi}{5},\frac{8\pi}{5}$ per orientation are required.

The need for a large number of orientations $\alpha$ is however relaxed due to the enlarged disks. Considering maximum speed we chose four orientations $\alpha =  0,\frac{1\pi}{4}, \frac{2\pi}{4},\frac{3\pi}{4}$ resulting in a total of 20 images to sufficiently cover the highly enlarged observable area. When speed is not the major goal one can chose more orientations $\alpha$ to obtain a higher quality what is also considered in regular SIM.

After obtaining the individual Fourier components $\tilde{n}(\mathbf{k})$, $\tilde{n}(\mathbf{k} \pm  \mathbf{k}_0)$ and $\tilde{n}(\mathbf{k} \pm 2  \mathbf{k}_0)$ they need to be assembled properly in reciprocal space, that is applying the same procedure which is conducted in SIM. The extracted raw components are so far scaled by the circularly symmetric OTF $H_2(\mathbf{k})$ or by its shifted versions $H_2(\mathbf{k} \pm \mathbf{k}_0)$ and $H_2(\mathbf{k} \pm 2 \mathbf{k}_0)$. To obtain an approximately homogeneous disk we divide the enlarged observable area (cf. Fig.\,\ref{fig:SIM+ABM}) into subregions and rescale the components by use of a Wiener filter $\tilde{n}_{new}(\mathbf{k})= \tilde{n}(\mathbf{k})/(H_2(\mathbf{k}) + \gamma)$, where the constant $\gamma > 0$ prevents division by zero. In general the modulus of $\gamma$ depends on the signal-to-noise ratio (SNR) a given measurement provides. After assembly we apply a triangular apodization, resembling the Fourier transform of an Airy disk, to the homogeneous disk to reduce ringing in the final image \cite{Gustafsson2005}. The final image is obtained by taking the modulus of the inverse Fourier transform of the assembled (and post-processed) disk in reciprocal space. Using more advanced deconvolution methods proposed and applied in SIM would result in an even further enhanced resolution \cite{Shaw2014,Stallinga2016}.

\section{Simulations}

For the simulations we chose masks with point-like emitters and calculated data as it would be detected by a CCD with discrete and finite pixels. Note that, here we are assuming perfect data, i.e. discrete intensity values matching theoretical calculations without noise. Experimental requirements to obtain the sought after SIQCM signals with preferably high SNR, i.e. sufficient statistics for the second and higher-order correlations, will be discussed later. For rescaling by use of the Wiener filter we used $\gamma = 0.05$. The pixel size and post-processed area where chosen in such a way that the offsets in reciprocal space approximately match integer numbers as we considered a real valued sinusoidal modulation. To remove the necessity for integer numbers in reciprocal space (challenging to realize in a real experiment) one can use a complex wave vector in real space \cite{Gustafsson2005}.

\begin{figure}[ht!]%
\centering
\includegraphics[width=0.95 \linewidth]{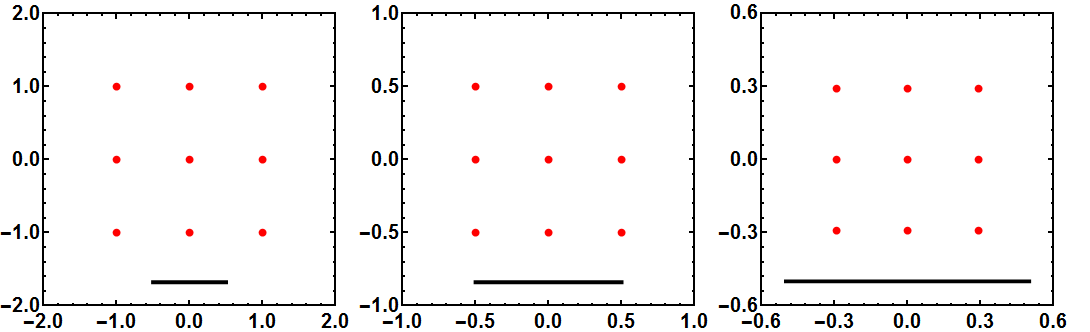}
\includegraphics[width=1.0 \linewidth]{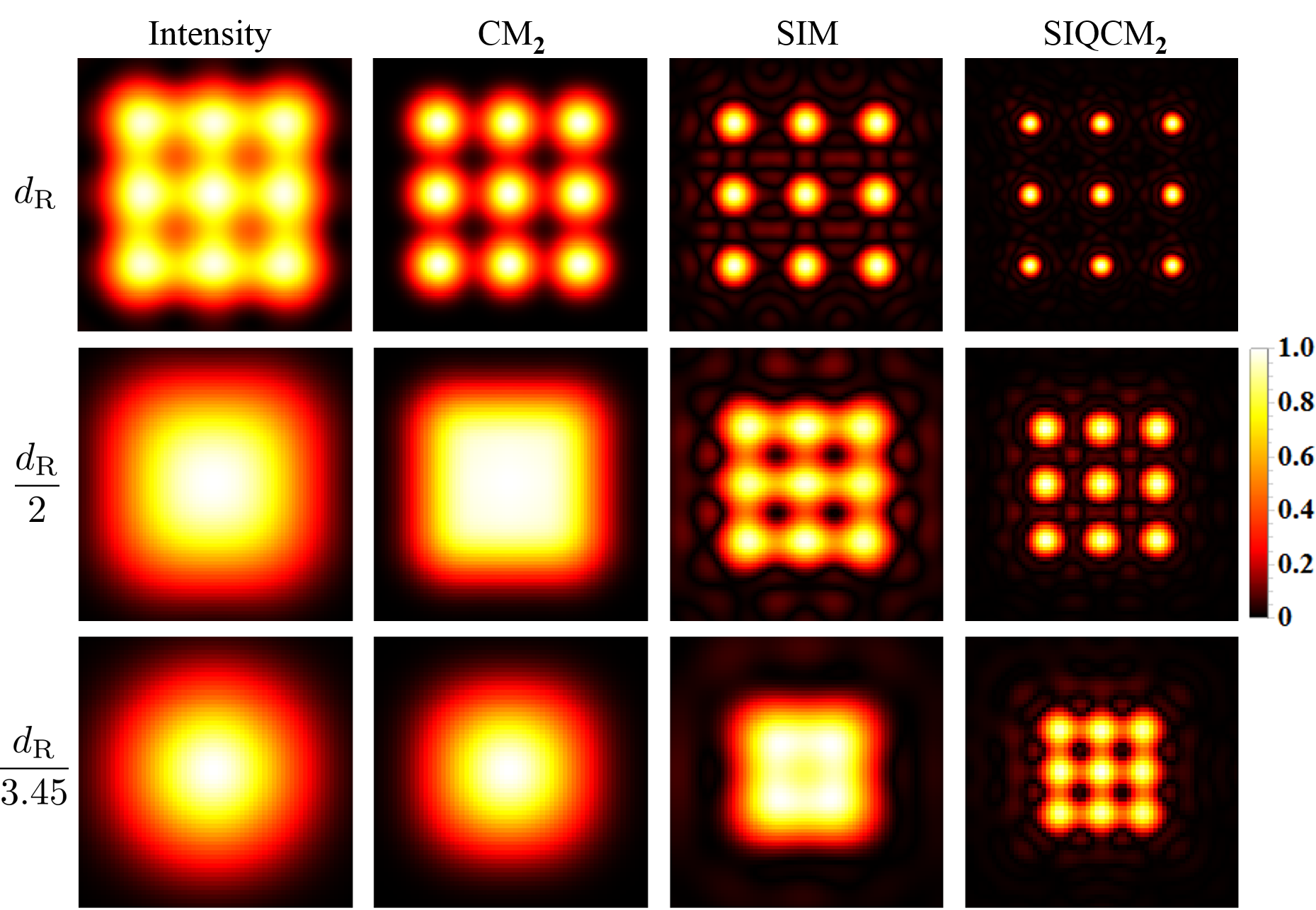}
\caption{Comparison of the resulting final images utilizing ordinary intensity measurements $G^{(1)}(\mathbf{r})$, $\text{CM}_2(\mathbf{r})$, SIM and $\text{SIQCM}_2(\mathbf{r})$  imaging a 3\,x\,3 array of independent emitters with separations $d=1.0\,d_\text{R}$, $d=0.5\,d_\text{R}$ and $d=0.29\,d_\text{R}$, see the masks at the top. The bar within each mask represents the Rayleigh limit $d_\text{R}$. The depicted areas in the final images differ from top to bottom as the sources are distributed over a smaller area. Though, the areas are not shrunk according to relative distances as the Airy disk's size in the intensity measurements $G^{(1)}(\mathbf{r})$ remains the same for each run.}
\label{fig:comparison}%
\end{figure}
\begin{figure}[ht!]%
\centering
\includegraphics[width=1.0 \linewidth]{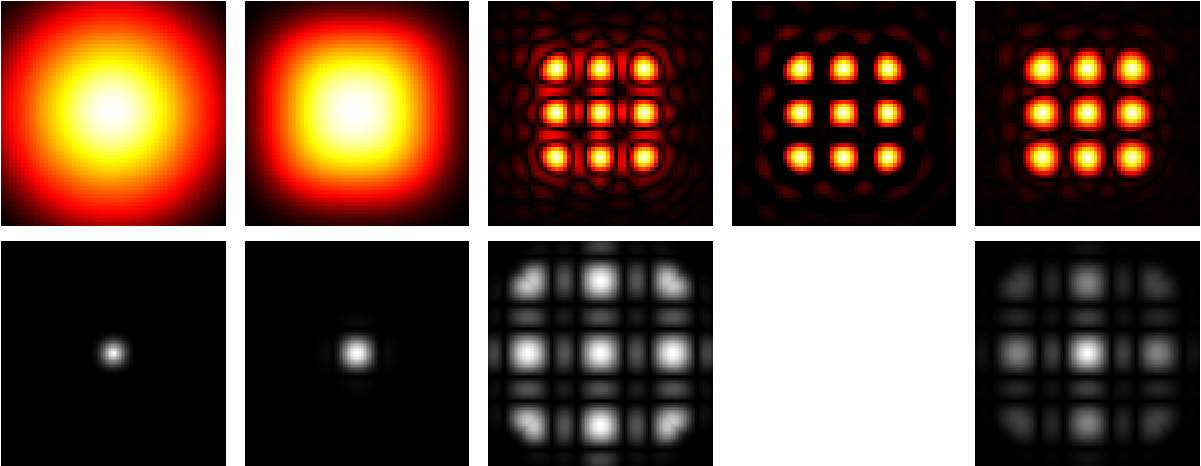}
\caption{Imaging a 3\,x\,3 array  with $d=0.29\,d_\text{R}$. Upper images show resulting distributions in real space and lower images show the corresponding reciprocal space. Images show from left to right: $G^{(1)}(\mathbf{r})$, $\text{CM}_3(\mathbf{r})$ and $\text{SIQCM}_3(\mathbf{r})$ with three different reconstruction approaches using a homogenous disk, a homogenous disk with cropped negative values and with a triangular apodization.}
\label{fig:G3SIM}%
\end{figure}
Simulations illustrating the resolution power of ordinary intensity measurements $G^{(1)}(\mathbf{r})$, $\text{CM}_2(\mathbf{r})$, SIM and $\text{SIQCM}_2(\mathbf{r})$ are presented in Fig.\,\ref{fig:comparison}, where a 3\,x\,3 array of independent emitters with separations $d=1.0\,d_\text{R}$, $d=0.5\,d_\text{R}$ and $d=0.29\,d_\text{R}$ is imaged by use of the enlisted techniques. The first array is resolved by every method, as the chosen distance corresponds to the classical resolution limit, however with $G^{(1)}(\mathbf{r})$ barely resolving individual emitters. $\text{CM}_2$ provides a moderately increased resolution and SIM the second best resolution power. We want to point out that even though  $\text{CM}_2$ and SIM already provide superresolved images, $\text{SIQCM}_2$ outperforms both methods by far and provides the highest resolution power. Reducing the source separation to $d=0.5\,d_\text{R}$ only SIM and $\text{SIQCM}_2$ can resolve the individual emitters and finally, for $d=0.29\,d_\text{R}$, only our new method resolves the array. Resolving the last array corresponds to a resolution improvement of 3.45, exactly matching the theoretical prediction.

To show the resolution power of our technique that scales very favorably as $m+\sqrt{m}$ compared to ordinary $\text{CM}$ that merely scales as $\sqrt{m}$ we also present simulations for third-order $\text{SIQCM}$ (see illustration on the right hand side in Fig.\,\ref{fig:SIM+ABM}). We chose six orientations $\alpha =  0,\frac{1\pi}{6}, \frac{2\pi}{6},\frac{3\pi}{6}, \frac{4\pi}{6},\frac{5\pi}{6}$ resulting in a total of 42 images as seven phases $\varphi$ are required per orientation. In Fig.\,\ref{fig:G3SIM} the resulting final images for the same 3\,x\,3 array with $d=0.29\,d_\text{R}$ imaged by use of $G^{(1)}(\mathbf{r})$, $\text{CM}_3(\mathbf{r})$ and with three different reconstruction approaches for $\text{SIQCM}_3(\mathbf{r})$ are presented together with the corresponding observable region in reciprocal space. The source distribution that was previously just resolved by $\text{SIQCM}_2$ is not resolved by $\text{CM}_3$ but clearly resolved by the $\text{SIQCM}_3$ signal. For the first reconstruction approach we subdivided the observable regions in sections and rescaled the Fourier components by the Wiener filter. The resulting image (third column) is simply the modulus of the Inverse Fourier transform of the homogenous disk. To remove the ringing we further omitted small imaginary parts acquired throughout the numerical evaluation (which should be zero in theory) and cropped negative values (fourth column). Note though that this approach might not be used this easily on real data with noise. The reconstruction method presented in the last column shows the triangular apodization applied to the homogenous disk. This method has also been used to produce the images in the last column of Fig.\,\ref{fig:comparison}. The final image is again simply the modulus of the Inverse Fourier transform. In the given simulation the approach that crops negative values to remove ringing performs best and provides the smallest FWHM of the effective PSF.

\begin{figure}[ht!]%
\centering
\includegraphics[width=0.95 \linewidth]{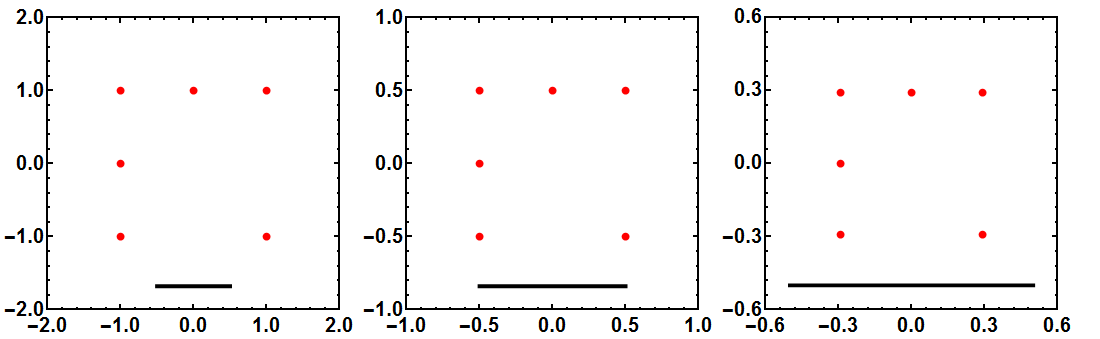}
\includegraphics[width=1.0 \linewidth]{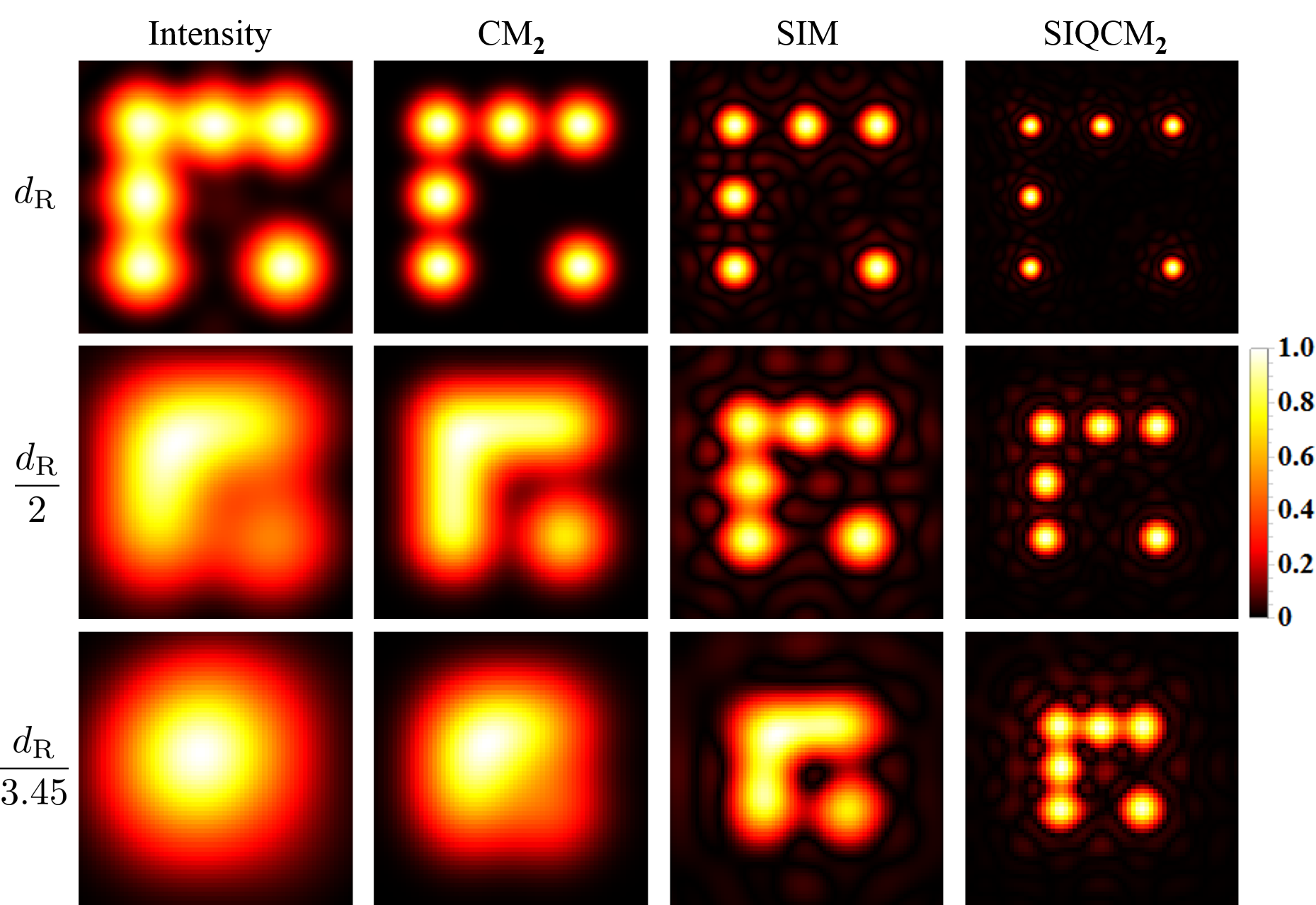}
\caption{Resulting final images utilizing ordinary intensity measurements $G^{(1)}(\mathbf{r})$, $\text{CM}_2(\mathbf{r})$, SIM and $\text{SIQCM}_2(\mathbf{r})$  imaging an irregular 3\,x\,3 array with three missing emitters with grid separations $d=1.0\,d_\text{R}$, $d=0.5\,d_\text{R}$ and $d=0.29\,d_\text{R}$, see the masks at the top. The bar within each mask represents the Rayleigh limit $d_\text{R}$.} 
\label{fig:comparison2}%
\end{figure}

\begin{figure}[ht!]%
\centering
\includegraphics[width=1.0 \linewidth]{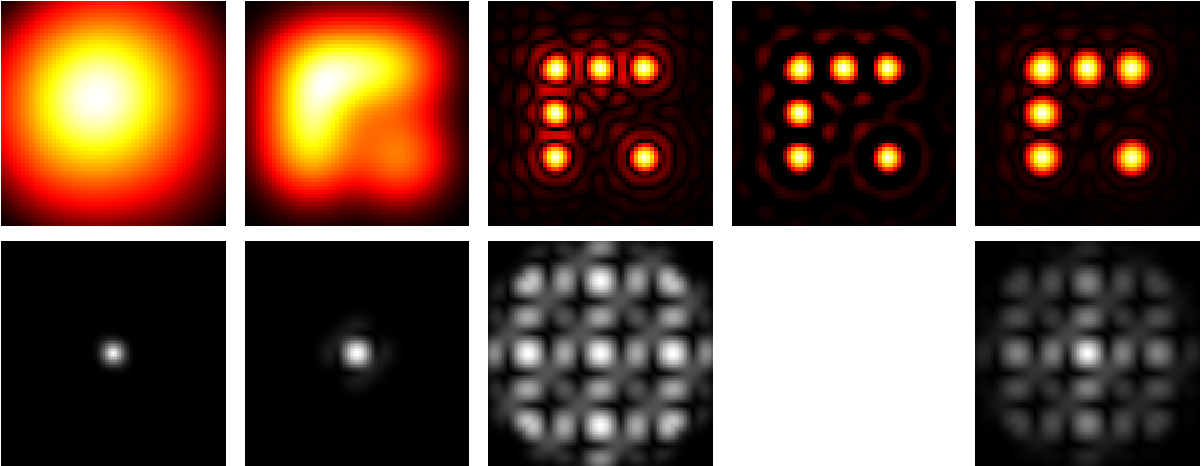}
\caption{Imaging an irregular 3\,x\,3 array  with three missing emitters with grid separation $d=0.29\,d_\text{R}$. Upper images show resulting distributions in real space and lower images show the corresponding reciprocal space. Images show from left to right: $G^{(1)}(\mathbf{r})$, $\text{CM}_3(\mathbf{r})$ and $\text{SIQCM}_3(\mathbf{r})$ with three different reconstruction approaches using a homogenous disk, a homogenous disk with cropped negative values and with a triangular apodization.}
\label{fig:G3SIM2}%
\end{figure}

Resulting images for another mask are depicted in Figs.\,\ref{fig:comparison2} and \ref{fig:G3SIM2} to show that the method can be applied to arbitrary emitter distributions. In contrast to the first mask three emitters have been omitted to obtain an irregular array.
The SIQCM signal again outperforms the CM and SIM signal by far, as was discussed in detail in the previous section.

\section{Conclusion and outlook}

We introduced a new quantum imaging technique we call SIQCM which is based on the profitable merger of linear SIM with anti-bunching CM. For a linear low-intensity standing wave illumination pattern and linear detection of photon auto correlations in the image plane of a microscope our technique provides in theory unlimited superresolution with improvement scaling favorably as $m +\sqrt{m}$ with the correlation order $m$. Hence, it has the potential to increase the spatial resolution in imaging a variety of samples and in particular biological ones. 
Further, we anticipate the SIQCM concept to be applicable to super-poissonian bunched light emission (e.g. used in SOFI, due to on-off blinking of fluorophores), where auto correlations in the image plane can be combined into cumulants that equally lead to a signal with narrowed PSF. Adding structured illumination will not only introduce offsets by $\pm \mathbf{k_0}$ but also higher harmonics with offsets up to $\pm m \mathbf{k_0}$.
Optical sectioning capability provided by CM as well as SIM can also be implemented enabling three-dimensional imaging with most likely increased axial resolution, due to higher harmonics in z-direction compared to 3D-SIM \cite{Gustafsson2008}. 

Our new SIQCM approach would bring similar benefits to two-photon microscopy \cite{Webb1990}, and vice versa. Considering a standing wave excitation pattern with wavelength within the red or near infrared part of the spectrum, short wavelength  photons from the UV or blue part of the spectrum are emitted by fluorophores due to two-photon absorption. Since the absorption cross section is inherently dependent on the squared excitation intensity the resulting effective illumination structure is of the form $I_{\text{str}}(\mathbf{r},t)= I_0 [  \frac{1}{2} + \frac{1}{2} \cos (\frac{\mathbf{k}_0}{2} \mathbf{r} + \varphi ) ]^2$, where $\frac{\mathbf{k}_0}{2} $ corresponds to the near infrared illumination wavelength and shifts by $\pm \mathbf{k}_0$ (corresponding to the fluorescence wavelength) as used in regular SIM already appear in the fluorescence intensity signal. Evaluating correlations additionally would result in taking the $2m$-th power of the structured illumination resulting in higher harmonics with offsets up to $\pm m \mathbf{k}_0$. The well-known advantages of two-photon microscopy, high penetration depth, energy deposition (and thus photobleaching) only within the vicinity of the focal plane and inherent optical sectioning capability would be added to our highly improved superresolution. 

Using bunched light emission our approach should be applicable with state of the art technology and reasonable speed as SOFI already provides acquisition times of a few seconds. To obtain the sought-after CM and SIQCM signals auto correlations in the image plane can also be determined by evaluating cross correlations of neighboring pixels, what reduces experimental requirements and introduces an effectively denser sampling in the image plane \cite{Schwartz2012,Schwartz2013}. The latter fact is of practical importance as resolutions achievable with SIQCM often exceed the sampling density of CCD cameras in use and thus interpolation can be circumvented.

\section*{Funding Information}

A.C. and J.v.Z. gratefully acknowledge funding by
the Erlangen Graduate School in Advanced Optical
Technologies (SAOT) by the German Research
Foundation (DFG) in the framework of the German
excellence initiative. G.S.A. thanks the BioPhotonics initiative of the Texas A\&M University for support. M.O.S. thanks the Robert A. Welch Foundation (Award Number: A-1261) and the Office of Naval Research (Award Number: \mbox{N00014-16-1-3054}) for support.

\section*{Acknowledgments}

A.C. gratefully acknowledges the hospitality of Texas A\&M University where this work was done.


\begin{thebibliography}{41}%
\makeatletter
\providecommand \@ifxundefined [1]{%
 \@ifx{#1\undefined}
}%
\providecommand \@ifnum [1]{%
 \ifnum #1\expandafter \@firstoftwo
 \else \expandafter \@secondoftwo
 \fi
}%
\providecommand \@ifx [1]{%
 \ifx #1\expandafter \@firstoftwo
 \else \expandafter \@secondoftwo
 \fi
}%
\providecommand \natexlab [1]{#1}%
\providecommand \enquote  [1]{``#1''}%
\providecommand \bibnamefont  [1]{#1}%
\providecommand \bibfnamefont [1]{#1}%
\providecommand \citenamefont [1]{#1}%
\providecommand \href@noop [0]{\@secondoftwo}%
\providecommand \href [0]{\begingroup \@sanitize@url \@href}%
\providecommand \@href[1]{\@@startlink{#1}\@@href}%
\providecommand \@@href[1]{\endgroup#1\@@endlink}%
\providecommand \@sanitize@url [0]{\catcode `\\12\catcode `\$12\catcode
  `\&12\catcode `\#12\catcode `\^12\catcode `\_12\catcode `\%12\relax}%
\providecommand \@@startlink[1]{}%
\providecommand \@@endlink[0]{}%
\providecommand \url  [0]{\begingroup\@sanitize@url \@url }%
\providecommand \@url [1]{\endgroup\@href {#1}{\urlprefix }}%
\providecommand \urlprefix  [0]{URL }%
\providecommand \Eprint [0]{\href }%
\providecommand \doibase [0]{http://dx.doi.org/}%
\providecommand \selectlanguage [0]{\@gobble}%
\providecommand \bibinfo  [0]{\@secondoftwo}%
\providecommand \bibfield  [0]{\@secondoftwo}%
\providecommand \translation [1]{[#1]}%
\providecommand \BibitemOpen [0]{}%
\providecommand \bibitemStop [0]{}%
\providecommand \bibitemNoStop [0]{.\EOS\space}%
\providecommand \EOS [0]{\spacefactor3000\relax}%
\providecommand \BibitemShut  [1]{\csname bibitem#1\endcsname}%
\let\auto@bib@innerbib\@empty
\bibitem [{\citenamefont {Abbe}(1873)}]{Abbe1873}%
  \BibitemOpen
  \bibfield  {author} {\bibinfo {author} {\bibfnamefont {E.}~\bibnamefont
  {Abbe}},\ }\href {\doibase doi:10.1007/BF02956173} {\bibfield  {journal}
  {\bibinfo  {journal} {Archiv f. mikrosk. Anatomie}\ }\textbf {\bibinfo
  {volume} {9}},\ \bibinfo {pages} {413} (\bibinfo {year} {1873})}\BibitemShut
  {NoStop}%
\bibitem [{\citenamefont {Rayleigh}(1879)}]{Rayleigh1879-2}%
  \BibitemOpen
  \bibfield  {author} {\bibinfo {author} {\bibfnamefont {L.}~\bibnamefont
  {Rayleigh}},\ }\href {\doibase 10.1080/14786447908639684} {\bibfield
  {journal} {\bibinfo  {journal} {Phil. Mag.}\ }\textbf {\bibinfo {volume}
  {8}},\ \bibinfo {pages} {261} (\bibinfo {year} {1879})}\BibitemShut {NoStop}%
\bibitem [{\citenamefont {Hell}\ and\ \citenamefont
  {Wichmann}(1994)}]{Hell1994}%
  \BibitemOpen
  \bibfield  {author} {\bibinfo {author} {\bibfnamefont {S.~W.}\ \bibnamefont
  {Hell}}\ and\ \bibinfo {author} {\bibfnamefont {J.}~\bibnamefont
  {Wichmann}},\ }\href {\doibase 10.1364/OL.19.000780} {\bibfield  {journal}
  {\bibinfo  {journal} {Opt. Lett.}\ }\textbf {\bibinfo {volume} {19}},\
  \bibinfo {pages} {780} (\bibinfo {year} {1994})}\BibitemShut {NoStop}%
\bibitem [{\citenamefont {Hell}(2015)}]{Hell2015}%
  \BibitemOpen
  \bibfield  {author} {\bibinfo {author} {\bibfnamefont {S.~W.}\ \bibnamefont
  {Hell}},\ }\href {\doibase 10.1002/anie.201504181} {\bibfield  {journal}
  {\bibinfo  {journal} {Angew. Chem. Int. Ed.}\ }\textbf {\bibinfo {volume}
  {54}},\ \bibinfo {pages} {8054} (\bibinfo {year} {2015})}\BibitemShut
  {NoStop}%
\bibitem [{\citenamefont {Hell}\ and\ \citenamefont {Kroug}(1995)}]{Hell1995}%
  \BibitemOpen
  \bibfield  {author} {\bibinfo {author} {\bibfnamefont {S.~W.}\ \bibnamefont
  {Hell}}\ and\ \bibinfo {author} {\bibfnamefont {M.}~\bibnamefont {Kroug}},\
  }\href {\doibase 10.1007/BF01081333} {\bibfield  {journal} {\bibinfo
  {journal} {Appl. Phys. B}\ }\textbf {\bibinfo {volume} {60}},\ \bibinfo
  {pages} {495} (\bibinfo {year} {1995})}\BibitemShut {NoStop}%
\bibitem [{\citenamefont {Hofmann}\ \emph {et~al.}(2005)\citenamefont
  {Hofmann}, \citenamefont {Eggeling}, \citenamefont {Jakobs},\ and\
  \citenamefont {Hell}}]{Hell2005}%
  \BibitemOpen
  \bibfield  {author} {\bibinfo {author} {\bibfnamefont {M.}~\bibnamefont
  {Hofmann}}, \bibinfo {author} {\bibfnamefont {C.}~\bibnamefont {Eggeling}},
  \bibinfo {author} {\bibfnamefont {S.}~\bibnamefont {Jakobs}}, \ and\ \bibinfo
  {author} {\bibfnamefont {S.~W.}\ \bibnamefont {Hell}},\ }\href {\doibase
  10.1073/pnas.0506010102} {\bibfield  {journal} {\bibinfo  {journal} {Proc.
  Natl. Acad. Sci. USA}\ }\textbf {\bibinfo {volume} {102}},\ \bibinfo {pages}
  {17565} (\bibinfo {year} {2005})}\BibitemShut {NoStop}%
\bibitem [{\citenamefont {Dickson}\ \emph {et~al.}(1997)\citenamefont
  {Dickson}, \citenamefont {Cubitt}, \citenamefont {Tsien},\ and\ \citenamefont
  {Moerner}}]{Moerner1997}%
  \BibitemOpen
  \bibfield  {author} {\bibinfo {author} {\bibfnamefont {R.~M.}\ \bibnamefont
  {Dickson}}, \bibinfo {author} {\bibfnamefont {A.~B.}\ \bibnamefont {Cubitt}},
  \bibinfo {author} {\bibfnamefont {R.~Y.}\ \bibnamefont {Tsien}}, \ and\
  \bibinfo {author} {\bibfnamefont {W.~E.}\ \bibnamefont {Moerner}},\ }\href
  {\doibase 10.1038/41048} {\bibfield  {journal} {\bibinfo  {journal} {Nature}\
  }\textbf {\bibinfo {volume} {388}},\ \bibinfo {pages} {355 } (\bibinfo {year}
  {1997})}\BibitemShut {NoStop}%
\bibitem [{\citenamefont {Hess}\ \emph {et~al.}(2006)\citenamefont {Hess},
  \citenamefont {Girirajan},\ and\ \citenamefont {Mason}}]{Mason2006}%
  \BibitemOpen
  \bibfield  {author} {\bibinfo {author} {\bibfnamefont {S.~T.}\ \bibnamefont
  {Hess}}, \bibinfo {author} {\bibfnamefont {T.~P.}\ \bibnamefont {Girirajan}},
  \ and\ \bibinfo {author} {\bibfnamefont {M.~D.}\ \bibnamefont {Mason}},\
  }\href {\doibase 10.1529/biophysj.106.091116} {\bibfield  {journal} {\bibinfo
   {journal} {Biophys. J.}\ }\textbf {\bibinfo {volume} {91}},\ \bibinfo
  {pages} {4258} (\bibinfo {year} {2006})}\BibitemShut {NoStop}%
\bibitem [{\citenamefont {Rust}\ \emph {et~al.}(2006)\citenamefont {Rust},
  \citenamefont {Bates},\ and\ \citenamefont {Zhuang}}]{Xiaowei2006}%
  \BibitemOpen
  \bibfield  {author} {\bibinfo {author} {\bibfnamefont {M.~J.}\ \bibnamefont
  {Rust}}, \bibinfo {author} {\bibfnamefont {M.}~\bibnamefont {Bates}}, \ and\
  \bibinfo {author} {\bibfnamefont {X.}~\bibnamefont {Zhuang}},\ }\href
  {\doibase 10.1038/nmeth929} {\bibfield  {journal} {\bibinfo  {journal} {Nat.
  Meth.}\ }\textbf {\bibinfo {volume} {3}},\ \bibinfo {pages} {793} (\bibinfo
  {year} {2006})}\BibitemShut {NoStop}%
\bibitem [{\citenamefont {Betzig}\ \emph {et~al.}(2006)\citenamefont {Betzig},
  \citenamefont {Patterson}, \citenamefont {Sougrat}, \citenamefont
  {Lindwasser}, \citenamefont {Olenych}, \citenamefont {Bonifacino},
  \citenamefont {Davidson}, \citenamefont {Lippincott-Schwartz},\ and\
  \citenamefont {Hess}}]{Betzig2006}%
  \BibitemOpen
  \bibfield  {author} {\bibinfo {author} {\bibfnamefont {E.}~\bibnamefont
  {Betzig}}, \bibinfo {author} {\bibfnamefont {G.~H.}\ \bibnamefont
  {Patterson}}, \bibinfo {author} {\bibfnamefont {R.}~\bibnamefont {Sougrat}},
  \bibinfo {author} {\bibfnamefont {O.~W.}\ \bibnamefont {Lindwasser}},
  \bibinfo {author} {\bibfnamefont {S.}~\bibnamefont {Olenych}}, \bibinfo
  {author} {\bibfnamefont {J.~S.}\ \bibnamefont {Bonifacino}}, \bibinfo
  {author} {\bibfnamefont {M.~W.}\ \bibnamefont {Davidson}}, \bibinfo {author}
  {\bibfnamefont {J.}~\bibnamefont {Lippincott-Schwartz}}, \ and\ \bibinfo
  {author} {\bibfnamefont {H.~F.}\ \bibnamefont {Hess}},\ }\href {\doibase
  10.1126/science.1127344} {\bibfield  {journal} {\bibinfo  {journal}
  {Science}\ }\textbf {\bibinfo {volume} {313}},\ \bibinfo {pages} {1642}
  (\bibinfo {year} {2006})}\BibitemShut {NoStop}%
\bibitem [{\citenamefont {Thiel}\ \emph {et~al.}(2007)\citenamefont {Thiel},
  \citenamefont {Bastin}, \citenamefont {Martin}, \citenamefont {Solano},
  \citenamefont {von Zanthier},\ and\ \citenamefont {Agarwal}}]{Thiel2007}%
  \BibitemOpen
  \bibfield  {author} {\bibinfo {author} {\bibfnamefont {C.}~\bibnamefont
  {Thiel}}, \bibinfo {author} {\bibfnamefont {T.}~\bibnamefont {Bastin}},
  \bibinfo {author} {\bibfnamefont {J.}~\bibnamefont {Martin}}, \bibinfo
  {author} {\bibfnamefont {E.}~\bibnamefont {Solano}}, \bibinfo {author}
  {\bibfnamefont {J.}~\bibnamefont {von Zanthier}}, \ and\ \bibinfo {author}
  {\bibfnamefont {G.~S.}\ \bibnamefont {Agarwal}},\ }\href {\doibase
  10.1103/PhysRevLett.99.133603} {\bibfield  {journal} {\bibinfo  {journal}
  {Phys. Rev. Lett.}\ }\textbf {\bibinfo {volume} {99}},\ \bibinfo {pages}
  {133603} (\bibinfo {year} {2007})}\BibitemShut {NoStop}%
\bibitem [{\citenamefont {Oppel}\ \emph {et~al.}(2012)\citenamefont {Oppel},
  \citenamefont {B\"uttner}, \citenamefont {Kok},\ and\ \citenamefont {von
  Zanthier}}]{Oppel2012a}%
  \BibitemOpen
  \bibfield  {author} {\bibinfo {author} {\bibfnamefont {S.}~\bibnamefont
  {Oppel}}, \bibinfo {author} {\bibfnamefont {T.}~\bibnamefont {B\"uttner}},
  \bibinfo {author} {\bibfnamefont {P.}~\bibnamefont {Kok}}, \ and\ \bibinfo
  {author} {\bibfnamefont {J.}~\bibnamefont {von Zanthier}},\ }\href {\doibase
  10.1103/PhysRevLett.109.233603} {\bibfield  {journal} {\bibinfo  {journal}
  {Phys. Rev. Lett.}\ }\textbf {\bibinfo {volume} {109}},\ \bibinfo {pages}
  {233603} (\bibinfo {year} {2012})}\BibitemShut {NoStop}%
\bibitem [{\citenamefont {Classen}\ \emph {et~al.}(2016)\citenamefont
  {Classen}, \citenamefont {Waldmann}, \citenamefont {Giebel}, \citenamefont
  {Schneider}, \citenamefont {Bhatti}, \citenamefont {Mehringer},\ and\
  \citenamefont {von Zanthier}}]{Classen2016}%
  \BibitemOpen
  \bibfield  {author} {\bibinfo {author} {\bibfnamefont {A.}~\bibnamefont
  {Classen}}, \bibinfo {author} {\bibfnamefont {F.}~\bibnamefont {Waldmann}},
  \bibinfo {author} {\bibfnamefont {S.}~\bibnamefont {Giebel}}, \bibinfo
  {author} {\bibfnamefont {R.}~\bibnamefont {Schneider}}, \bibinfo {author}
  {\bibfnamefont {D.}~\bibnamefont {Bhatti}}, \bibinfo {author} {\bibfnamefont
  {T.}~\bibnamefont {Mehringer}}, \ and\ \bibinfo {author} {\bibfnamefont
  {J.}~\bibnamefont {von Zanthier}},\ }\href {\doibase
  10.1103/PhysRevLett.117.253601} {\bibfield  {journal} {\bibinfo  {journal}
  {Phys. Rev. Lett.}\ }\textbf {\bibinfo {volume} {117}},\ \bibinfo {pages}
  {253601} (\bibinfo {year} {2016})}\BibitemShut {NoStop}%
\bibitem [{\citenamefont {Dertinger}\ \emph {et~al.}(2009)\citenamefont
  {Dertinger}, \citenamefont {Colyer}, \citenamefont {Iyer}, \citenamefont
  {Weiss},\ and\ \citenamefont {Enderlein}}]{Dertinger2009}%
  \BibitemOpen
  \bibfield  {author} {\bibinfo {author} {\bibfnamefont {T.}~\bibnamefont
  {Dertinger}}, \bibinfo {author} {\bibfnamefont {R.}~\bibnamefont {Colyer}},
  \bibinfo {author} {\bibfnamefont {G.}~\bibnamefont {Iyer}}, \bibinfo {author}
  {\bibfnamefont {S.}~\bibnamefont {Weiss}}, \ and\ \bibinfo {author}
  {\bibfnamefont {J.}~\bibnamefont {Enderlein}},\ }\href {\doibase
  10.1073/pnas.0907866106} {\bibfield  {journal} {\bibinfo  {journal} {Proc.
  Natl. Acad. Sci. USA}\ }\textbf {\bibinfo {volume} {106}},\ \bibinfo {pages}
  {22287} (\bibinfo {year} {2009})}\BibitemShut {NoStop}%
\bibitem [{\citenamefont {Schwartz}\ and\ \citenamefont
  {Oron}(2012)}]{Schwartz2012}%
  \BibitemOpen
  \bibfield  {author} {\bibinfo {author} {\bibfnamefont {O.}~\bibnamefont
  {Schwartz}}\ and\ \bibinfo {author} {\bibfnamefont {D.}~\bibnamefont
  {Oron}},\ }\href {\doibase 10.1103/PhysRevA.85.033812} {\bibfield  {journal}
  {\bibinfo  {journal} {Phys. Rev. A}\ }\textbf {\bibinfo {volume} {85}},\
  \bibinfo {pages} {033812} (\bibinfo {year} {2012})}\BibitemShut {NoStop}%
\bibitem [{\citenamefont {Schwartz}\ \emph {et~al.}(2013)\citenamefont
  {Schwartz}, \citenamefont {Levitt}, \citenamefont {Tenne}, \citenamefont
  {Itzhakov}, \citenamefont {Deutsch},\ and\ \citenamefont
  {Oron}}]{Schwartz2013}%
  \BibitemOpen
  \bibfield  {author} {\bibinfo {author} {\bibfnamefont {O.}~\bibnamefont
  {Schwartz}}, \bibinfo {author} {\bibfnamefont {J.~M.}\ \bibnamefont
  {Levitt}}, \bibinfo {author} {\bibfnamefont {R.}~\bibnamefont {Tenne}},
  \bibinfo {author} {\bibfnamefont {S.}~\bibnamefont {Itzhakov}}, \bibinfo
  {author} {\bibfnamefont {Z.}~\bibnamefont {Deutsch}}, \ and\ \bibinfo
  {author} {\bibfnamefont {D.}~\bibnamefont {Oron}},\ }\href {\doibase
  10.1021/nl402552m} {\bibfield  {journal} {\bibinfo  {journal} {Nano Lett.}\
  }\textbf {\bibinfo {volume} {13}},\ \bibinfo {pages} {5832} (\bibinfo {year}
  {2013})}\BibitemShut {NoStop}%
\bibitem [{\citenamefont {Gatto~Monticone}\ \emph {et~al.}(2014)\citenamefont
  {Gatto~Monticone}, \citenamefont {Katamadze}, \citenamefont {Traina},
  \citenamefont {Moreva}, \citenamefont {Forneris}, \citenamefont
  {Ruo-Berchera}, \citenamefont {Olivero}, \citenamefont {Degiovanni},
  \citenamefont {Brida},\ and\ \citenamefont {Genovese}}]{Genovese2014a}%
  \BibitemOpen
  \bibfield  {author} {\bibinfo {author} {\bibfnamefont {D.}~\bibnamefont
  {Gatto~Monticone}}, \bibinfo {author} {\bibfnamefont {K.}~\bibnamefont
  {Katamadze}}, \bibinfo {author} {\bibfnamefont {P.}~\bibnamefont {Traina}},
  \bibinfo {author} {\bibfnamefont {E.}~\bibnamefont {Moreva}}, \bibinfo
  {author} {\bibfnamefont {J.}~\bibnamefont {Forneris}}, \bibinfo {author}
  {\bibfnamefont {I.}~\bibnamefont {Ruo-Berchera}}, \bibinfo {author}
  {\bibfnamefont {P.}~\bibnamefont {Olivero}}, \bibinfo {author} {\bibfnamefont
  {I.~P.}\ \bibnamefont {Degiovanni}}, \bibinfo {author} {\bibfnamefont
  {G.}~\bibnamefont {Brida}}, \ and\ \bibinfo {author} {\bibfnamefont
  {M.}~\bibnamefont {Genovese}},\ }\href {\doibase
  10.1103/PhysRevLett.113.143602} {\bibfield  {journal} {\bibinfo  {journal}
  {Phys. Rev. Lett.}\ }\textbf {\bibinfo {volume} {113}},\ \bibinfo {pages}
  {143602} (\bibinfo {year} {2014})}\BibitemShut {NoStop}%
\bibitem [{\citenamefont {Gustafsson}(2000)}]{Gustafsson2000}%
  \BibitemOpen
  \bibfield  {author} {\bibinfo {author} {\bibfnamefont {M.~G.~L.}\
  \bibnamefont {Gustafsson}},\ }\href {\doibase
  10.1046/j.1365-2818.2000.00710.x} {\bibfield  {journal} {\bibinfo  {journal}
  {J. Microsc.}\ }\textbf {\bibinfo {volume} {198}},\ \bibinfo {pages} {82}
  (\bibinfo {year} {2000})}\BibitemShut {NoStop}%
\bibitem [{\citenamefont {Gustafsson}(2005)}]{Gustafsson2005}%
  \BibitemOpen
  \bibfield  {author} {\bibinfo {author} {\bibfnamefont {M.~G.~L.}\
  \bibnamefont {Gustafsson}},\ }\href {\doibase 10.1073/pnas.0406877102}
  {\bibfield  {journal} {\bibinfo  {journal} {Proc. Natl. Acad. Sci. USA}\
  }\textbf {\bibinfo {volume} {102}},\ \bibinfo {pages} {13081} (\bibinfo
  {year} {2005})}\BibitemShut {NoStop}%
\bibitem [{\citenamefont {Hajek}\ \emph {et~al.}(2010)\citenamefont {Hajek},
  \citenamefont {Littleton}, \citenamefont {Turk}, \citenamefont {McIntyre},\
  and\ \citenamefont {Rubinsztein-Dunlop}}]{Rubinsztein2010}%
  \BibitemOpen
  \bibfield  {author} {\bibinfo {author} {\bibfnamefont {K.~M.}\ \bibnamefont
  {Hajek}}, \bibinfo {author} {\bibfnamefont {B.}~\bibnamefont {Littleton}},
  \bibinfo {author} {\bibfnamefont {D.}~\bibnamefont {Turk}}, \bibinfo {author}
  {\bibfnamefont {T.~J.}\ \bibnamefont {McIntyre}}, \ and\ \bibinfo {author}
  {\bibfnamefont {H.}~\bibnamefont {Rubinsztein-Dunlop}},\ }\href {\doibase
  10.1364/OE.18.019263} {\bibfield  {journal} {\bibinfo  {journal} {Opt.
  Express}\ }\textbf {\bibinfo {volume} {18}},\ \bibinfo {pages} {19263}
  (\bibinfo {year} {2010})}\BibitemShut {NoStop}%
\bibitem [{\citenamefont {Park}\ \emph {et~al.}(2014)\citenamefont {Park},
  \citenamefont {Lee}, \citenamefont {Lee},\ and\ \citenamefont
  {Lee}}]{Lee2014}%
  \BibitemOpen
  \bibfield  {author} {\bibinfo {author} {\bibfnamefont {J.~H.}\ \bibnamefont
  {Park}}, \bibinfo {author} {\bibfnamefont {S.-W.}\ \bibnamefont {Lee}},
  \bibinfo {author} {\bibfnamefont {E.~S.}\ \bibnamefont {Lee}}, \ and\
  \bibinfo {author} {\bibfnamefont {J.~Y.}\ \bibnamefont {Lee}},\ }\href
  {\doibase 10.1364/OE.22.009854} {\bibfield  {journal} {\bibinfo  {journal}
  {Opt. Express}\ }\textbf {\bibinfo {volume} {22}},\ \bibinfo {pages} {9854}
  (\bibinfo {year} {2014})}\BibitemShut {NoStop}%
\bibitem [{\citenamefont {Zeng}\ \emph {et~al.}(2014)\citenamefont {Zeng},
  \citenamefont {Al-Amri},\ and\ \citenamefont {Zubairy}}]{Zubairy2014}%
  \BibitemOpen
  \bibfield  {author} {\bibinfo {author} {\bibfnamefont {X.}~\bibnamefont
  {Zeng}}, \bibinfo {author} {\bibfnamefont {M.}~\bibnamefont {Al-Amri}}, \
  and\ \bibinfo {author} {\bibfnamefont {M.~S.}\ \bibnamefont {Zubairy}},\
  }\href {\doibase 10.1103/PhysRevB.90.235418} {\bibfield  {journal} {\bibinfo
  {journal} {Phys. Rev. B}\ }\textbf {\bibinfo {volume} {90}},\ \bibinfo
  {pages} {235418} (\bibinfo {year} {2014})}\BibitemShut {NoStop}%
\bibitem [{\citenamefont {Agarwal}\ and\ \citenamefont
  {Kapale}(2006)}]{Agarwal2006a}%
  \BibitemOpen
  \bibfield  {author} {\bibinfo {author} {\bibfnamefont {G.~S.}\ \bibnamefont
  {Agarwal}}\ and\ \bibinfo {author} {\bibfnamefont {K.~T.}\ \bibnamefont
  {Kapale}},\ }\href {\doibase 10.1088/0953-4075/39/17/002} {\bibfield
  {journal} {\bibinfo  {journal} {J. Phys. B}\ }\textbf {\bibinfo {volume}
  {39}},\ \bibinfo {pages} {3437} (\bibinfo {year} {2006})}\BibitemShut
  {NoStop}%
\bibitem [{\citenamefont {Miles}\ \emph {et~al.}(2013)\citenamefont {Miles},
  \citenamefont {Simmons},\ and\ \citenamefont {Yavuz}}]{Yavuz2013}%
  \BibitemOpen
  \bibfield  {author} {\bibinfo {author} {\bibfnamefont {J.~A.}\ \bibnamefont
  {Miles}}, \bibinfo {author} {\bibfnamefont {Z.~J.}\ \bibnamefont {Simmons}},
  \ and\ \bibinfo {author} {\bibfnamefont {D.~D.}\ \bibnamefont {Yavuz}},\
  }\href {\doibase 10.1103/PhysRevX.3.031014} {\bibfield  {journal} {\bibinfo
  {journal} {Phys. Rev. X}\ }\textbf {\bibinfo {volume} {3}},\ \bibinfo {pages}
  {031014} (\bibinfo {year} {2013})}\BibitemShut {NoStop}%
\bibitem [{\citenamefont {Miles}\ \emph {et~al.}(2015)\citenamefont {Miles},
  \citenamefont {Das}, \citenamefont {Simmons},\ and\ \citenamefont
  {Yavuz}}]{Yavuz2015}%
  \BibitemOpen
  \bibfield  {author} {\bibinfo {author} {\bibfnamefont {J.~A.}\ \bibnamefont
  {Miles}}, \bibinfo {author} {\bibfnamefont {D.}~\bibnamefont {Das}}, \bibinfo
  {author} {\bibfnamefont {Z.~J.}\ \bibnamefont {Simmons}}, \ and\ \bibinfo
  {author} {\bibfnamefont {D.~D.}\ \bibnamefont {Yavuz}},\ }\href {\doibase
  10.1103/PhysRevA.92.033838} {\bibfield  {journal} {\bibinfo  {journal} {Phys.
  Rev. A}\ }\textbf {\bibinfo {volume} {92}},\ \bibinfo {pages} {033838}
  (\bibinfo {year} {2015})}\BibitemShut {NoStop}%
\bibitem [{\citenamefont {Kiffner}\ \emph {et~al.}(2008)\citenamefont
  {Kiffner}, \citenamefont {Evers},\ and\ \citenamefont
  {Zubairy}}]{Zubairy2008}%
  \BibitemOpen
  \bibfield  {author} {\bibinfo {author} {\bibfnamefont {M.}~\bibnamefont
  {Kiffner}}, \bibinfo {author} {\bibfnamefont {J.}~\bibnamefont {Evers}}, \
  and\ \bibinfo {author} {\bibfnamefont {M.~S.}\ \bibnamefont {Zubairy}},\
  }\href {\doibase 10.1103/PhysRevLett.100.073602} {\bibfield  {journal}
  {\bibinfo  {journal} {Phys. Rev. Lett.}\ }\textbf {\bibinfo {volume} {100}},\
  \bibinfo {pages} {073602} (\bibinfo {year} {2008})}\BibitemShut {NoStop}%
\bibitem [{\citenamefont {Liao}\ \emph {et~al.}(2010)\citenamefont {Liao},
  \citenamefont {Al-Amri},\ and\ \citenamefont {Zubairy}}]{Zubairy2010}%
  \BibitemOpen
  \bibfield  {author} {\bibinfo {author} {\bibfnamefont {Z.}~\bibnamefont
  {Liao}}, \bibinfo {author} {\bibfnamefont {M.}~\bibnamefont {Al-Amri}}, \
  and\ \bibinfo {author} {\bibfnamefont {M.~S.}\ \bibnamefont {Zubairy}},\
  }\href {\doibase 10.1103/PhysRevLett.105.183601} {\bibfield  {journal}
  {\bibinfo  {journal} {Phys. Rev. Lett.}\ }\textbf {\bibinfo {volume} {105}},\
  \bibinfo {pages} {183601} (\bibinfo {year} {2010})}\BibitemShut {NoStop}%
\bibitem [{\citenamefont {Kapale}\ and\ \citenamefont
  {Agarwal}(2010)}]{Kapale2010}%
  \BibitemOpen
  \bibfield  {author} {\bibinfo {author} {\bibfnamefont {K.~T.}\ \bibnamefont
  {Kapale}}\ and\ \bibinfo {author} {\bibfnamefont {G.~S.}\ \bibnamefont
  {Agarwal}},\ }\href {\doibase 10.1364/OL.35.002792} {\bibfield  {journal}
  {\bibinfo  {journal} {Opt. Lett.}\ }\textbf {\bibinfo {volume} {35}},\
  \bibinfo {pages} {2792} (\bibinfo {year} {2010})}\BibitemShut {NoStop}%
\bibitem [{\citenamefont {Dertinger}\ \emph {et~al.}(2012)\citenamefont
  {Dertinger}, \citenamefont {Xu}, \citenamefont {Naini}, \citenamefont
  {Vogel},\ and\ \citenamefont {Weiss}}]{Dertinger2012}%
  \BibitemOpen
  \bibfield  {author} {\bibinfo {author} {\bibfnamefont {T.}~\bibnamefont
  {Dertinger}}, \bibinfo {author} {\bibfnamefont {J.}~\bibnamefont {Xu}},
  \bibinfo {author} {\bibfnamefont {O.~F.}\ \bibnamefont {Naini}}, \bibinfo
  {author} {\bibfnamefont {R.}~\bibnamefont {Vogel}}, \ and\ \bibinfo {author}
  {\bibfnamefont {S.}~\bibnamefont {Weiss}},\ }\href {\doibase
  10.1186/2192-2853-1-2} {\bibfield  {journal} {\bibinfo  {journal} {Opt.
  Nanoscopy}\ }\textbf {\bibinfo {volume} {1}},\ \bibinfo {pages} {2} (\bibinfo
  {year} {2012})}\BibitemShut {NoStop}%
\bibitem [{\citenamefont {Gustafsson}\ \emph {et~al.}(2008)\citenamefont
  {Gustafsson}, \citenamefont {Shao}, \citenamefont {Carlton}, \citenamefont
  {Wang}, \citenamefont {Golubovskaya}, \citenamefont {Cande}, \citenamefont
  {Agard},\ and\ \citenamefont {Sedat}}]{Gustafsson2008}%
  \BibitemOpen
  \bibfield  {author} {\bibinfo {author} {\bibfnamefont {M.~G.~L.}\
  \bibnamefont {Gustafsson}}, \bibinfo {author} {\bibfnamefont
  {L.}~\bibnamefont {Shao}}, \bibinfo {author} {\bibfnamefont {P.~M.}\
  \bibnamefont {Carlton}}, \bibinfo {author} {\bibfnamefont {C.~J.~R.}\
  \bibnamefont {Wang}}, \bibinfo {author} {\bibfnamefont {I.~N.}\ \bibnamefont
  {Golubovskaya}}, \bibinfo {author} {\bibfnamefont {W.~Z.}\ \bibnamefont
  {Cande}}, \bibinfo {author} {\bibfnamefont {D.~A.}\ \bibnamefont {Agard}}, \
  and\ \bibinfo {author} {\bibfnamefont {J.~W.}\ \bibnamefont {Sedat}},\ }\href
  {\doibase 10.1529/biophysj.107.120345} {\bibfield  {journal} {\bibinfo
  {journal} {Biophys. J.}\ }\textbf {\bibinfo {volume} {94}},\ \bibinfo {pages}
  {4957 } (\bibinfo {year} {2008})}\BibitemShut {NoStop}%
\bibitem [{\citenamefont {Basch\'e}\ \emph {et~al.}(1992)\citenamefont
  {Basch\'e}, \citenamefont {Moerner}, \citenamefont {Orrit},\ and\
  \citenamefont {Talon}}]{Basche1992}%
  \BibitemOpen
  \bibfield  {author} {\bibinfo {author} {\bibfnamefont {T.}~\bibnamefont
  {Basch\'e}}, \bibinfo {author} {\bibfnamefont {W.~E.}\ \bibnamefont
  {Moerner}}, \bibinfo {author} {\bibfnamefont {M.}~\bibnamefont {Orrit}}, \
  and\ \bibinfo {author} {\bibfnamefont {H.}~\bibnamefont {Talon}},\ }\href
  {\doibase 10.1103/PhysRevLett.69.1516} {\bibfield  {journal} {\bibinfo
  {journal} {Phys. Rev. Lett.}\ }\textbf {\bibinfo {volume} {69}},\ \bibinfo
  {pages} {1516} (\bibinfo {year} {1992})}\BibitemShut {NoStop}%
\bibitem [{\citenamefont {Lounis}\ \emph {et~al.}(2000)\citenamefont {Lounis},
  \citenamefont {Bechtel}, \citenamefont {Gerion}, \citenamefont {Alivisatos},\
  and\ \citenamefont {Moerner}}]{Moerner2000}%
  \BibitemOpen
  \bibfield  {author} {\bibinfo {author} {\bibfnamefont {B.}~\bibnamefont
  {Lounis}}, \bibinfo {author} {\bibfnamefont {H.}~\bibnamefont {Bechtel}},
  \bibinfo {author} {\bibfnamefont {D.}~\bibnamefont {Gerion}}, \bibinfo
  {author} {\bibfnamefont {P.}~\bibnamefont {Alivisatos}}, \ and\ \bibinfo
  {author} {\bibfnamefont {W.}~\bibnamefont {Moerner}},\ }\href {\doibase
  10.1016/S0009-2614(00)01042-3} {\bibfield  {journal} {\bibinfo  {journal}
  {Chem. Phys. Lett.}\ }\textbf {\bibinfo {volume} {329}},\ \bibinfo {pages}
  {399 } (\bibinfo {year} {2000})}\BibitemShut {NoStop}%
\bibitem [{\citenamefont {Brouri}\ \emph {et~al.}(2000)\citenamefont {Brouri},
  \citenamefont {Beveratos}, \citenamefont {Poizat},\ and\ \citenamefont
  {Grangier}}]{Grangier2000}%
  \BibitemOpen
  \bibfield  {author} {\bibinfo {author} {\bibfnamefont {R.}~\bibnamefont
  {Brouri}}, \bibinfo {author} {\bibfnamefont {A.}~\bibnamefont {Beveratos}},
  \bibinfo {author} {\bibfnamefont {J.-P.}\ \bibnamefont {Poizat}}, \ and\
  \bibinfo {author} {\bibfnamefont {P.}~\bibnamefont {Grangier}},\ }\href
  {\doibase 10.1364/OL.25.001294} {\bibfield  {journal} {\bibinfo  {journal}
  {Opt. Lett.}\ }\textbf {\bibinfo {volume} {25}},\ \bibinfo {pages} {1294}
  (\bibinfo {year} {2000})}\BibitemShut {NoStop}%
\bibitem [{\citenamefont {Michler}\ \emph {et~al.}(2000)\citenamefont
  {Michler}, \citenamefont {Imamoglu}, \citenamefont {Mason}, \citenamefont
  {Carson}, \citenamefont {Strouse},\ and\ \citenamefont
  {Buratto}}]{Buratto2000}%
  \BibitemOpen
  \bibfield  {author} {\bibinfo {author} {\bibfnamefont {P.}~\bibnamefont
  {Michler}}, \bibinfo {author} {\bibfnamefont {A.}~\bibnamefont {Imamoglu}},
  \bibinfo {author} {\bibfnamefont {M.~D.}\ \bibnamefont {Mason}}, \bibinfo
  {author} {\bibfnamefont {P.~J.}\ \bibnamefont {Carson}}, \bibinfo {author}
  {\bibfnamefont {G.~F.}\ \bibnamefont {Strouse}}, \ and\ \bibinfo {author}
  {\bibfnamefont {S.~K.}\ \bibnamefont {Buratto}},\ }\href {\doibase
  10.1038/35023100} {\bibfield  {journal} {\bibinfo  {journal} {Nature}\
  }\textbf {\bibinfo {volume} {406}},\ \bibinfo {pages} {968} (\bibinfo {year}
  {2000})}\BibitemShut {NoStop}%
\bibitem [{\citenamefont {Born}\ and\ \citenamefont
  {Wolf}(1999)}]{BornWolf1999}%
  \BibitemOpen
  \bibfield  {author} {\bibinfo {author} {\bibfnamefont {M.}~\bibnamefont
  {Born}}\ and\ \bibinfo {author} {\bibfnamefont {E.}~\bibnamefont {Wolf}},\
  }\href@noop {} {\emph {\bibinfo {title} {Principles of Optics}}},\ \bibinfo
  {edition} {7th}\ ed.\ (\bibinfo  {publisher} {Cambridge University Press},\
  \bibinfo {year} {1999})\BibitemShut {NoStop}%
\bibitem [{\citenamefont {Agarwal}(2012)}]{Agarwal2012}%
  \BibitemOpen
  \bibfield  {author} {\bibinfo {author} {\bibfnamefont {G.~S.}\ \bibnamefont
  {Agarwal}},\ }\href@noop {} {\emph {\bibinfo {title} {Quantum Optics}}}\
  (\bibinfo  {publisher} {Cambridge University Press},\ \bibinfo {year}
  {2012})\BibitemShut {NoStop}%
\bibitem [{\citenamefont {Geissbuehler}\ \emph {et~al.}(2012)\citenamefont
  {Geissbuehler}, \citenamefont {Bocchio}, \citenamefont {Dellagiacoma},
  \citenamefont {Berclaz}, \citenamefont {Leutenegger},\ and\ \citenamefont
  {Lasser}}]{Geissbuehler2012}%
  \BibitemOpen
  \bibfield  {author} {\bibinfo {author} {\bibfnamefont {S.}~\bibnamefont
  {Geissbuehler}}, \bibinfo {author} {\bibfnamefont {N.~L.}\ \bibnamefont
  {Bocchio}}, \bibinfo {author} {\bibfnamefont {C.}~\bibnamefont
  {Dellagiacoma}}, \bibinfo {author} {\bibfnamefont {C.}~\bibnamefont
  {Berclaz}}, \bibinfo {author} {\bibfnamefont {M.}~\bibnamefont
  {Leutenegger}}, \ and\ \bibinfo {author} {\bibfnamefont {T.}~\bibnamefont
  {Lasser}},\ }\href {\doibase 10.1186/2192-2853-1-4} {\bibfield  {journal}
  {\bibinfo  {journal} {Opt. Nanoscopy}\ }\textbf {\bibinfo {volume} {1}},\
  \bibinfo {pages} {4} (\bibinfo {year} {2012})}\BibitemShut {NoStop}%
\bibitem [{\citenamefont {Glauber}(1963)}]{Glauber1963-2}%
  \BibitemOpen
  \bibfield  {author} {\bibinfo {author} {\bibfnamefont {R.~J.}\ \bibnamefont
  {Glauber}},\ }\href {\doibase 10.1103/PhysRev.130.2529} {\bibfield  {journal}
  {\bibinfo  {journal} {Phys. Rev.}\ }\textbf {\bibinfo {volume} {130}},\
  \bibinfo {pages} {2529} (\bibinfo {year} {1963})}\BibitemShut {NoStop}%
\bibitem [{\citenamefont {O'Holleran}\ and\ \citenamefont
  {Shaw}(2014)}]{Shaw2014}%
  \BibitemOpen
  \bibfield  {author} {\bibinfo {author} {\bibfnamefont {K.}~\bibnamefont
  {O'Holleran}}\ and\ \bibinfo {author} {\bibfnamefont {M.}~\bibnamefont
  {Shaw}},\ }\href {\doibase 10.1364/BOE.5.002580} {\bibfield  {journal}
  {\bibinfo  {journal} {Biomed. Opt. Express}\ }\textbf {\bibinfo {volume}
  {5}},\ \bibinfo {pages} {2580} (\bibinfo {year} {2014})}\BibitemShut
  {NoStop}%
\bibitem [{\citenamefont {Chakrova}\ \emph {et~al.}(2016)\citenamefont
  {Chakrova}, \citenamefont {Rieger},\ and\ \citenamefont
  {Stallinga}}]{Stallinga2016}%
  \BibitemOpen
  \bibfield  {author} {\bibinfo {author} {\bibfnamefont {N.}~\bibnamefont
  {Chakrova}}, \bibinfo {author} {\bibfnamefont {B.}~\bibnamefont {Rieger}}, \
  and\ \bibinfo {author} {\bibfnamefont {S.}~\bibnamefont {Stallinga}},\ }\href
  {\doibase 10.1364/JOSAA.33.000B12} {\bibfield  {journal} {\bibinfo  {journal}
  {J. Opt. Soc. Am. A}\ }\textbf {\bibinfo {volume} {33}},\ \bibinfo {pages}
  {B12} (\bibinfo {year} {2016})}\BibitemShut {NoStop}%
\bibitem [{\citenamefont {Denk}\ \emph {et~al.}(1990)\citenamefont {Denk},
  \citenamefont {Strickler},\ and\ \citenamefont {Webb}}]{Webb1990}%
  \BibitemOpen
  \bibfield  {author} {\bibinfo {author} {\bibfnamefont {W.}~\bibnamefont
  {Denk}}, \bibinfo {author} {\bibfnamefont {J.~H.}\ \bibnamefont {Strickler}},
  \ and\ \bibinfo {author} {\bibfnamefont {W.~W.}\ \bibnamefont {Webb}},\
  }\href {\doibase 10.1126/science.2321027} {\bibfield  {journal} {\bibinfo
  {journal} {Science}\ }\textbf {\bibinfo {volume} {248}},\ \bibinfo {pages}
  {73} (\bibinfo {year} {1990})}\BibitemShut {NoStop}%
\end{thebibliography}

%

\end{document}